
\input jnl
\def\ifundefined#1{\expandafter\ifx\csname#1\endcsname\relax}
\ifundefined{equationfile}\input eqnorder
\fi
\ifundefined{referencefile}\input reforder
\fi\gdef\refto#1{ [#1]}\citeall\refto
\catcode`@=11
\def\call#1{$\rm\each@rg\callr@nge{#1}$}
\catcode`@=12
\oneandathirdspace

{\singlespace\rightline  {NSF-ITP-94-119}\rightline  {gr-qc/9412066}}

\title
	Universal Scaling and Echoing in Gravitational Collapse of a
	Complex Scalar Field
\author
	Eric W. Hirschmann\footnote{$^{\rm a)}$}%
	{Author's electronic address: \tt ehirsch@dolphin.physics.ucsb.edu\hfil}
\affil
	Department of Physics
	University of California
	Santa Barbara, CA 93106-9530
\author
	Douglas M. Eardley\footnote{$^{\rm b)}$}%
	{Author's electronic address: \tt doug@itp.ucsb.edu\hfil}
\affil
	Institute for Theoretical Physics
	University of California
	Santa Barbara, CA 93106-4030
\received
\abstract{\oneandathirdspace
This paper studies gravitational collapse of a complex scalar
field at the threshold for black hole formation, assuming that the
collapse is spherically symmetric and continuously self-similar.  A new
solution of the coupled Einstein-scalar field equations is derived, after
a small amount of numerical work with ordinary differential equations.
The universal scaling and echoing behavior discovered by Choptuik in
spherically symmetrical gravitational collapse appear in a somewhat
different form.  Properties of the endstate of the collapse are derived:
The collapse leaves behind an irregular outgoing pulse of scalar
radiation, with exactly flat spacetime within it.}

\body
\head{I. Introduction}

Recently there has been a lot of new interest in gravitational collapse
just at the threshold for formation of black holes, inspired by the
striking numerical results of Choptuik\refto{Chop} on spherically
symmetric gravitational collapse of scalar field configurations.  Further
numerical results for vacuum relativity in axial symmetry by Abrahams and
Evans\refto{AE} suggest that the phenomena discovered by Choptuik not just
restricted to spherical symmetry, and a number of extensions have been
proposed\refto{Brady, ONT, ST, JT, DG}.

Gravitational collapse has two kinds of possible endstates, according to
current views.  The first endstate consists of a black hole, plus some
outgoing matter and outgoing gravitational radiation.  The second endstate
consists of a stationary remnant star, plus some outgoing matter and
outgoing gravitational radiation, but no black hole.  Choice between these
endstates depends on initial conditions; gravitational collapse ends in a
black hole only if the gravitational field becomes strong enough, in the
sense of the singularity theorems\refto{HE}.

A third possibility, the naked singularity, is thought not to occur
(cosmic censorship conjecture\refto{PenCC}) but is not rigorously ruled
out.

The thought experiment employed by Choptuik, in the context of numerical
relativity, is to "tune" across the critical threshold in the space of
initial conditions that separates the non-black-hole endstate from the
black hole endstate, and to carefully study the critical behavior of
various quantities at this threshold\refto{Chop,DC}.  Among the infinite
number of parameters that characterize the initial state, he chooses to
tune a single convenient parameter $p$ that influences the strength of
gravity in the initial state.  One could imagine that the threshold
behavior would be intricate, and depend strongly on exactly how one tunes
across the threshold.  However, he finds impressive evidence that the
threshold behavior is, in important respects, universal.

\def\Mbh{M_{\rm BH}}
\def\iw{{i\omega}}
\def\Eq#1{Eq.~\(#1)}
\def\Eqs#1{Eqs.~\(#1)}

\def\Equations#1{Equations~\(#1)}

For gravitational collapse of fermionic matter from suitably regular
initial conditions, the Chandrasekhar mass sets the scale of minimum mass
for any black hole, as long as the initial conditions are sufficiently
regular.  Some condition on regularity of the initial conditions is
clearly necessary;  otherwise one could create black holes of arbitrarily
small mass with focused beams of ultrarelativistic fermions, evading the
threshold set by the Chandrasekhar limit.  Barring irregular initial
conditions, the black hole mass $\Mbh(p)$ for fermionic configurations
therefore behaves in a simple and rather uninteresting way at threshold:
It is a step function of $p$.

For gravitational collapse of bosonic matter from regular initial
conditions, one might have thought that $\Mbh(p)$ would likewise be a step
function of $p$, with the role of the absent Chandrasekhar mass being
played by some mass-scale of the initial data.  However, this is not the
case.  In spherically symmetric collapse of a massless real scalar field
$\phi$ coupled to gravity, Choptuik finds power-law behavior
$$
  \Mbh(p) \propto (p-p^*)^\gamma, \quad \gamma\approx0.37 \eqno(mbh)
$$
at threshold, and conjectures that $\gamma$ might be universal.  For a
massless scalar field with no self-coupling, no bound star can exist, so
that the endstate consists of outgoing scalar radiation, plus a possible
black hole.

Furthermore, exactly at the threshold $p^*$, he finds a unique field
configuration acting as an attractor for all nearby initial conditions.
This field configuration --- which we will call a ``choptuon" --- has a
discrete self-similarity, by virtue of which it exhibits a striking,
recurrent ``echoing" behavior:  it repeats itself at ever-decreasing time-
and length-scales
$$\eqalignno{
	t'	&= e^{-n\Delta}t		&(dss a)	\cr
	r'	&= e^{-n\Delta}r		&(dss b)	\cr
	{ds'}^2	&= e^{-2n\Delta}ds^2		&(dss c)	\cr
   \phi(t',r')	&= \phi(t,r)			&(dss d)	\cr
		&(n = 1,2,3,\ldots)		&(dss e)	\cr
}$$
where $(t,r)$ are spherical coordinates, and $\Delta\approx30$ is a
constant belonging to the choptuon, determined numerically\refto{Chop}.
The solution is thus invariant under a discrete family of scale
transformations.  This choptuon is itself regular, and acts as an
attractor for a very wide class of regular initial data in spherical
symmetry.

Why study the field configuration exactly at threshold?  After all, we
understand what happens just below threshold (outgoing radiation) and just
above (tiny black hole plus outgoing radiation).  Beyond the obvious
fascination of the choptuons, there are at least two important
motivations:  First, choptuons amount to a new kind of counterexample to
some formulations of the cosmic censorship conjecture, a counterexample
that cannot be blamed on bad choice of matter fields.  Technically they
are a counterexample because regions of arbitrarily strong curvature are
visible to observers at future null infinity.  One can always bypass this
new counterexample by reformulating the conjecture --- for instance, to
specify that {\it generically} such behavior does not occur --- but the
more serious point is that choptuons threaten to obstruct any proof of the
cosmic censorship conjecture using the global theory of nonlinear partial
differential equations.  Therefore it will be necessary to confront and
understand them.

Secondly, and even more importantly, choptuons represent in principle a
means by which effects of extremely strong field gravity, even quantum
gravity, can be observable in the present universe.  An experimenter,
tuning across the black hole threshold in a succession of gravitational
collapses, should be able observe events in which some outgoing radiation
appears from regions where the spacetime curvature is as strong as the
Planck value, and where quantum behavior of general relativity or string
theory may be studied.  Therefore, the quantum corrections to the
classical choptuon --- or, indeed, the quantum gravitational and stringy
generalizations of the choptuon --- demand study.

Striking though they are, the numerical computations involving partial
differential equations (PDE) do not give complete information about the
choptuons.  For instance, output is restricted to the domain covered by
the numerical coordinate system adopted, which may not cover the whole
domain of dependence of the initial data.  This makes it hard to study the
endstate, the spacetime singularity to which the choptuon collapses, and
the burst of radiation that it emits.  Therefore it is valuable to find
further choptuons that can be studied by analytic techniques, or
mostly-analytic techniques including numerical solution of ordinary
differential equations (ODE).  In particular, Evans and Coleman\refto{EC}
obtained choptuons for the spherically symmetric collapse of hot gas,
which are continuously self-similar, a special case of discretely
self-similar choptuons.  The Evans and Coleman configurations show
threshold behavior of $\Mbh$, but not echoing.  Continuously self-similar
solutions are convenient because they are governed by ODE, not PDE{\null}.
The greater numerical accuracy obtainable with ODEs is desirable, not for
its own sake, but because of the light it may shed on the deeper questions
of universality.

The subject of this paper is the behavior at threshold in gravitational
collapse of a complex scalar field $\phi$, under spherical symmetry.  The
use of a complex scalar field, rather than a real one, will allow
``echoing" to occur in the form of phase oscillations:  $\phi$ changes
phase, but not amplitude, under a scale transformation.  We thereby
construct and study a continuously self-similar, complex choptuon as
solutions of the coupled Einstein-scalar field equations.  In spherical
coordinates $(t,r)$ our scalar field takes the form
$$
        \phi(t,r) = (-t)^{\iw} f(-r/t),		\eqno(ssphi)
$$
where $\omega$ is a constant to be fixed by a nonlinear eigenvalue problem
arising from the field equations.  This field is continuously
self-similar:  under a scale transformation by an arbitrary parameter
$\lambda$, it transforms solely by a phase factor:
$$\eqalignno{
	t'	&= e^{-\lambda}t		&(css a)	\cr
	r'	&= e^{-\lambda}r		&(css b)	\cr
	{ds'}^2	&= e^{-2\lambda}ds^2		&(css c)	\cr
  \phi(t',r')	&= e^{-\iw\lambda}\phi(t,r)	&(css d)	\cr
	 	&(0<\lambda<\infty)		&(css e)	\cr
}$$
This solution thus exhibits ``linear" phase oscillations --- a form of the
``echoing" ---  superimposed on an exact continuous scale symmetry.  It is
thus similar to, but ``less nonlinear" than the solution of
Choptuik\refto{Chop}.

We are able to construct a choptuon as a nonextendible spacetime, singular
only at the single point $(t,r)=(0,0)$, that evolves from regular initial
data.  A burst of outgoing scalar radiation, irregular but of finite
energy, emerges just prior the retarded time at which the singular point
forms at $(0,0)$, and explodes along the future light cone of that point.
The most surprising aspect of this choptuon is that spacetime appears to
be precisely flat throughout the interior of the future light cone of the
singular point.  Thus, in this example anyway, a choptuon leaves behind
exactly flat space, with no radiation at all after the irregular outgoing
burst.

Since a real $\phi$ is a special case of a complex $\phi$, all of
Choptuik's numerical results immediately apply as special cases to a
complex field; in addition, we find exist intrinsically complex choptuons
which do not reduce to the real case.  Which of these choptuons is the
strongest attractor is an important question not addressed in this paper;
we will return to it in future papers.  Preliminary numerical evidence
by Choptuik\refto{cpriv} seems to show that the real choptuon is at least
not rendered unstable when complex degrees of freedom are added.

\head{II. Field Equations}

\let\D=\nabla
\let\al=\alpha
\def\dt{\partial_t}
\def\dr{\partial_r}
\def\dz#1{d#1\over dz}
\def\1#1{{1\over#1}}
\def\Re{\mathop{\rm Re}}
\def\bp{\beta_{\mathord+}}
\def\bm{\beta_{\mathord-}}
\def\phb{\phantom{-}b}
\def\phv{\phantom{-}v}
\def\L{{\cal L}}

We begin with the spacetime metric in the form used by Choptuik,\refto{Chop}
$$
	ds^2 = -\alpha^2dt^2 + a^2dr^2 +r^2d\Omega^2	\eqno(metric)
$$
where $\alpha(t,r)$ and $a(t,r)$ are functions of time and radius.
This is an example of ``radial gauge" because the area of spheres
(given by the coefficient of $d\Omega^2$) defines the radial coordinate.
Radial gauge breaks down at an apparent horizon, and so another coordinate
system will need to be used if an apparent horizon appears.  The time
coordinate is chosen so that gravitational collapse on the axis of
spherical symmetry first occurs
at $t=0$, and the metric is regular for $t<0$.  This metric
remains invariant in form under transformations
$$
	t=t(t') \eqno(ttran)
$$
of the time coordinate, and such a transformation can be used to set
$\alpha(t,0)=1$ for $t<0$ on the axis.  By regularity (no cone singularity),
$a(t,0)=1$ on the axis for $t<0$ as well.

Matter consists of a free, massless, complex scalar field $\phi$
that obeys the wave equation
$$
	\square\phi = 0
$$
while the Einstein equations are
$$\eqalignno{
 R_{\mu\nu} - \12g{\mu\nu}R	&= 8\pi T_{\mu\nu}\cr
				&= 8\pi\left(\D_\mu\phi \D_\nu\phi -
			\12 g_{\mu\nu}\D^\rho\phi\D_\rho\phi\right).\cr
}$$

These equations amount to
$$\eqalignno{
 \1{r^2}\dr\left({{\al\over a}r^2\dr\phi}\right)
			&= \dt\left({{a\over\al}\dt\phi}\right)	&(fe a)\cr
 \1\al\dr\al - \1a\dr a	&= {a^2-1\over r}			&(fe b)\cr
 {2\over a^2r}\left(\1\al\dr\al + \1a\dr a\right) &= 8\pi\left(\1{a^2}
	\dr\phi^*\dr\phi + \1{\al^2}\dt\phi\dt\phi^*\right)	&(fe c)\cr
 {2\over ar}\dt a	&= 4\pi\left(\dr\phi^*\dt\phi +
			\dr\phi\dt\phi^*\right)			&(fe d)\cr
}$$
These equations admit global $U(1)$ symmetries for a constant $\Lambda$,
$$
	\phi' = e^{i\Lambda}\phi, \quad -\infty<\Lambda<\infty \eqno(u1)
$$
leaving the metric invariant.

\head{III. Continuous Self-Similarity}

We now derive the form of the scaling transformations, \Eqs{css}, for
the fields.  We assume spacetime is spherically symmetric, and admits a
homothetic Killing vector field\refto{DE} $\xi$ obeying
$$
	\L_\xi g_{\mu\nu} =
	\D_\mu \xi_\nu+\D_\nu \xi_\mu = 2g_{\mu\nu},		\eqno(Lxi)
$$
which generates a continuous one parameter family of homothetic motions
(self-similarities) on spacetime, \Eqs{css}.  Here, $\L$ denotes the
Lie derivative.
In a coordinate system, \Eq{metric}, $\xi$ may always be taken in the form
$$
	\xi^\mu\partial_\mu = t\partial_t + r\partial_r,	\eqno(defxi)
$$
possibly after a further coordinate transformation, \Eq{ttran}.
\Equations{Lxi,defxi} are implemented by \Eqs{css abc}.  The metric
functions are then of the special form
$$
	\al(t,r) = \al(-r/t), \quad a(t,r) = a(-r/t).		\eqno(aal)
$$
The minus sign is chosen so that $-r/t>0$ where $t<0$ to the past of the
singularity.  Dimensional analysis (in classical general relativity, not
quantum field theory) suggests that $\phi(t,r)$ should have dimensions
(length)$^0$.  Therefore, should we assume that $\phi$ is invariant under
$\xi$, $\L_\xi\phi=0$ or $\phi(t,r)=\phi(r/t)$?  No, the global $U(1)$
symmetry must be accommodated too.  Quite generally, one must allow for
spacetime symmetries to get mixed up with internal symmetries;  for
instance, the Maxwell equations for the potential $A_\mu$ are conformally
invariant only in the sense that a gauge transformation be allowed to
accompany each conformal transformation.  In this case, we must allow some
$U(1)$ transformation, \Eq{u1}, to accompany each scale transformation,
$$
  \L_\xi\phi	= \xi^\mu\partial_\mu\phi = \iw\phi	\eqno(Lphi)
$$
under an infinitesimal scale transformation, or
$$
  \phi(t',r')	= \exp(-\iw\lambda)\phi(t,r)
$$
under the finite scale transformation, \Eqs{css}, with $\omega$ a constant
of the solution.

This transformation law for $\phi$ can be conveniently implemented
by adopting the form
$$
        \phi(t,r) = (-t)^{\iw} f(z).			\eqno(ssphiz)
$$
The time coordinate $t$ has now been redefined, \Eq{ttran}, so that the
first singularity of the collapse is at $t=0$; and a new independent
variable $z$ has been introduced by
$$
	z = - r/t;					\eqno(defz)
$$
$z$ is invariant under scale transformations, \Eqs{css}.  Also, from \Eq{aal},
$$
	\al = \al(z), a = a(z).				\eqno(aalz)
$$

With the scale symmetry implemented by \Eqs{ssphiz,defz,aalz}, the
next step is to solve the field equations.  We can give initial conditions
in the hypersurface $t=-1$ and then evolve forward in time.  It follows
from a theorem of Berger\refto{Bev} that the Einstein equations are
compatible with homothetic symmetry, in the sense that scale invariant
initial data will always evolve to a scale invariant spacetime;  therefore
the field equations ought to have a solution.

Remarks should be made as to how suitable the assumption of
self-similarity is.  Self-similar spacetimes can never be asymptotically
flat (because the ADM or Bondi mass would define a length scale, breaking
self-similarity); furthermore, there are some reasons to believe that
self-similar spacetimes can never be spatially compact\refto{EIMM}.
Spatial compactness is irrelevant here, but shouldn't gravitational
collapse be modelled by an asymptotically flat spacetime?  However, the
self-similar solution should be interpreted as a model of the
gravitationally collapsing region out to some radius $R$, where it can
match smoothly onto a non-self-similar, asymptotically flat region.  To
anticipate, we will find that our solutions always have a certain horizon
called the past similarity horizon; as long as the matching radius $R$ is
outside the past similarity horizon, the gravitational collapse that we
study here will remain entirely within the domain of dependence of the
self-similar region of the initial data.  For instance, for the complex
choptuon discussed below, a value $R>5.004$ suffices.

Transform the metric variables $\al$, $a$ to a new set $b$, $u$ given by
$$\eqalignno{
  b(z)	&= \al(z)/a(z),\cr
  u(z)	&= a^2(z)-1.\cr
}$$
Under the similarity hypothesis, $b=b(z)$ and $u=u(z)$ are functions
of $z\equiv -r/t$ alone.  Under this transformation, the metric, \Eq{metric},
becomes
$$
  ds^2 = e^{2\tau}\left( (1+u)\left[-(b^2-z^2)d\tau^2 +
	2d\tau dz+dz^2\right] + z^2d\Omega^2 \right)	\eqno(metricz)
$$
where $\tau\equiv\ln t$.  Following Choptuik's notation,
represent $\phi$ in terms of complex functions $(\Phi(t,r),\Pi(t,r))$ where
$$\eqalignno{
  \Phi	&= \dr\phi = (-t)^{\iw-1}q(z)\cr
  \Pi	&= {a\over\al}\dt\phi = -(-t)^{\iw-1}p(z)\cr
\noalign{where the functions $q(z)$,$p(z)$ are:}
  q(z)	&= {df\over dz}\cr
  p(z)	&= {a\over \al}\left(\iw f-z{df\over dz}\right)\cr
}$$

The field equations now take the form
$$\eqalignno{
  {d\over dz}\pmatrix{q\cr p}	&= -\1z\pmatrix{u+2&0\cr0&u} + \1\Delta\pmatrix
    {-z&\phb\cr\phb&-z}\pmatrix{\bp&0\cr0&\bm}\pmatrix{q\cr p},&(emz a)\cr
\dz b	&= {bu\over z},&(emz b)\cr
\dz u	&= (u+1)\left[4\pi z(\left|q\right|^2 + \left|p\right|^2) -
		{u\over z}\right],&(emz c)\cr
\dz u	&= -8\pi(u+1)b\Re(q^*p),&(emz d)\cr
}$$
where
$$\eqalignno{
\beta_{\mathord\pm}	&= \iw + u \pm 1,\cr
\Delta			&= b^2-z^2.\cr
}$$
The boundary conditions are as follows.  At $z=0$, regularity of solutions
on the axis of spherical symmetry demands
$$\eqalignno{
	b(0) &= 1,\cr
	u(0) &= 0,\cr
	q(0) &= 0.\cr
}$$
The boundary value $P=p(0)$ of $p$ is a free boundary condition, and
will henceforth be taken real and nonnegative by a global phase
transformation of $\phi$, without loss of generality.  Integration
of \Eqs{emz} from $z=0$ to $z=+\infty$ is tantamount to solving the
initial value problem in the spacelike hypersurface $t=-1$.  Evolution
in $t$ is then fixed by self-similarity.

The wave equation \Eq{emz a} has a singular point when $\Delta$ vanishes,
\ie when $b(z)=z$.  This value of $z$ will be called $z_2$,
and the point will be called a {\sl similarity horizon}.
It represents the null hypersurface in spacetime where the homothetic
Killing vector, timelike near the axis, becomes null.  The value of
$u$ there,
$$
\kappa\equiv1-u(z_2),
$$
is free, while $b(z_2)=z_2$ is of course determined by $z_2$.

The flat space wave equation provides a simple introduction;
see Fig.~1.{\null}  Here the metric is
$$
	ds^2 = -dt^2 + dr^2 +r^2 d\Omega^2	\eqno(fmetric)
$$
and the general spherically symmetric solution to the massless
wave equation is
$$
	\phi(t,r) = {f(t-r) + g(t+r)\over r}
$$
where $f$ and $g$ are arbitrary functions.  The homothetic killing
vector is $\xi=t\dt+r\dr$ and all self-similar solutions, \Eq{ssphi},
are linear combinations of
$$
  \phi^\pm(t,r) = {(t\mp r)^{\iw+1} \over r} =
			(-t)^\iw {(1\pm z)^{\iw+1}\over z} \eqno(phipm)
$$
where
$$
z=-r/t.
$$
Here $\phi^+$ is an outgoing wave, is regular at
the past similarity horizon $z=+1$, and is irregular at the
future similarity horizon $z=-1$.  In contrast, $\phi^-$ is an ingoing
wave, is irregular at the past similarity horizon $z=+1$, and is regular
at the future similarity horizon $z=-1$.  All solutions are regular
at the hypersurface $t=0$ or $z=\infty$, except at the single spacetime
point $(t,r)=(0,0)$.  Only the linear combination $\phi^+-\phi^-$ is
regular at the origin $r=0$ of spherical coordinates (for all $t\ne0$).
We can transform from the coordinates $(t,r)$ to the coordinates
$(\tau,z)$ where $\tau=\ln t$.  Then the metric becomes
$$
  ds^2 = e^{2\tau}\left( -(1-z^2)d\tau^2 +
	2d\tau dz+dz^2 + z^2d\Omega^2 \right)		\eqno(fmetricz)
$$
The coordinates $(\tau,z)$ have a singularity on the spacelike hyperplane
$t=0$.  Yet another coordinate system for flat spacetime, regular at $t=0$,
is $(\rho,v)$, where $\rho=\ln(r)$ and $v=1/z=-t/r$;  the metric is
$$
  ds^2 = e^{2\rho}\left( -dv^2 -
	2dv d\rho+(1-v^2)d\rho^2 + d\Omega^2 \right).	\eqno(fmetricv)
$$

Returning now to the curved spacetime metric \Eq{metric}, analysis of the
wave equation near $z=z_2$ shows that $q(z)$ and $p(z)$ likewise have a
regular solution and an irregular solution at the similarity horizon.  The
regular solution behaves like
$$\eqalignno{
	f	&\sim \const				&(reg a)\cr
    (q,p)	&\sim \const				&(reg b)\cr
\noalign{\noindent and the irregular solution behaves like}
	f	&\sim (z-z_2)^{(\iw+1)/\kappa}		&(irreg a)\cr
	(q,p)	&\sim (z-z_2)^{(\iw+1)/\kappa-1}	&(irreg b)\cr
}$$
This behavior is identical to the case of flat spacetime except for
one aspect:  The presence of the quantity
$$
	\kappa \equiv 1-{db\over dz}\Big|_{z=z_2}
$$
in the exponent.  For flat spacetime $\kappa=1$.  In a self-similar
spacetime, $\kappa$ plays a role parallel to that of the surface gravity
of the event horizon of a black hole in a stationary spacetime, and in
this paper we will simply call it the ``surface gravity".

The regular solution again represents outgoing radiation crossing the
similarity horizon, and the irregular solution represents ingoing
radiation propagating along the horizon.  Since the subject of study is
gravitational collapse from regular initial conditions, and since the
similarity horizon $z=z_2$ is in the Cauchy development of the initial
data --- it is to the past of the earliest singular point (t,r)=(0,0) ---
regularity of $\phi$ will be demanded on the similarity horizon.  This
means that a linear combination of $q$ and $p$ must vanish at $z=z_2$:
$$
	\bp q(z_2) - \bm p(z_2) = 0,		\eqno(bc)
$$
so that, for instance, we can choose $p(z_2)$ freely as a complex valued
boundary condition at the similarity horizon, and $q(z_2)$ is then fixed.

To sum up, the free data for the system will be taken as
$$\eqalignno{
	&\omega,					&(data a)\cr
	&P\equiv p(0) \quad{\rm(real)},			&(data b)\cr
	&z_2,						&(data c)\cr
	&\kappa=1-u(z_2),				&(data d)\cr
	&p(z_2) \quad{\rm(complex)},			&(data e)\cr
}$$
amounting to six real constants.  Solutions to the system \Eq{emz}
must be found by searching in the six-dimensional data space.

\head{IV.  Construction and Properties of the Complex Choptuon}

The numerical methods we used followed Numerical Recipes\refto{NR} Chapter
16: We handled the system as a two-point boundary value problem with one
fixed boundary (at $z=0$) and one free boundary (at $z=z_2$).  We ``shot"
with an adaptive step ODE solver from each boundary point, to meet at a
point $z_1$ in the middle.  The six free data values were adjusted in an
outer loop with a Newton's-method solver for nonlinear equations, the six
nonlinear equations being the matching conditions for the six ODEs \(emz)
at $z_1$.  Convergence of the Newton's method then identified a solution
on the domain $0\le z\le z_2$.  Initial work was done with a Runge-Kutta
integrator, while a Bulirsch-Stoer integrator was subsequently used for
higher precision.  The solution was then continued to larger $z$ without
the need for further boundary conditions.

To enforce numerically the boundary condition on the horizon is difficult,
because the irregular part of $\phi$, which we want to annul, vanishes
anyway if $\kappa<1$, and the smaller $\kappa$, the faster it vanishes.
Said differently, the difficulty is that we want to start a purely regular
solution at the horizon and continue it, but the unwanted irregular
solution grows rapidly as it is continued away from the solution;  this is
a numerically unstable situation.  We handled this by computing
analytically the second derivative of the regular solution at $z=z_2$, and
using a second-order Taylor expansion to start the solution there.  Even
so we encountered numerical problems for $\kappa\ltwid 1/3$, as discussed
in Appendix A.

A single new solution was found on a domain $0\le z<z_2$, obtained
with the values
$$\eqalignno{
	&\omega=1.9154446 \pm 0.0000001,\cr
	&P\equiv p(0)=0.67217263 \pm 0.0000004,\cr
	&z_2=5.0035380 \pm 0.0000002,\cr
	&u(z_2)\equiv1-\kappa=0.39707205 \pm 0.00000003,\cr
	&p(z_2)=(0.020305344 \pm 0.000000003) +
			(0.007153158 \pm 0.000000002)i,\cr
}$$
This solution will be called the ``complex choptuon."  Its complex
conjugate solution, with the sign of $\omega$ changed, also of course
exists.  Figure 2 displays the functions $q(z)$, $p(z))$,
$b(z)$, $u(z)$ for the complex choptuon.

We also found solution at $\omega=0$, corresponding to a real scalar
field $\phi$;  this, however, is nothing but the flat ($k=0$)
Robertson-Walker cosmological model with a real scalar field as matter,
reversed in time.  Further details of the numerical results are reported
in Appendix A.

This solution was continued smoothly past $z=z_2$ to large $z$.  The
nature of the point $z=\infty$ on the z-axis must now be clarified.
Examination shows this to be a singular point of \Eqs{emz}.
However, on physical grounds there should be no
spacetime singularity here, since $z=\infty$ corresponds to the spacelike
hypersurface $t=0$, which should be regular except at the axis, since it
lies in the Cauchy development of the initial data.  Any apparent
singularity there must fall under strong suspicion as a coordinate
singularity caused by a bad choice of time coordinate, \Eq{ttran}, near
$t=0$.  Indeed, the system of equations can be rendered regular at
$z=\infty$ by the following change of variables:
$$\eqalignno{
  dw	&= b(z){dz\over z^2}, \quad w=0\rm{\quad at\quad}z=\infty,&(new a)\cr
  \pmatrix{Q(w)\cr P(w)} &= z^{1-\iw}\pmatrix{q(z)\cr p(z)},&(new b)\cr
   v(w)	&= {b(z)\over z},&(new c)\cr
   u(w) &= u(z).&(new d)\cr
}$$
The spacetime metric, \Eqs{metric,metricz}, becomes
$$
  ds^2 = e^{2\rho}\left( (1+u)\left[-dw^2 +
	2dw d\rho+(1-v^2)d\rho^2\right] + d\Omega^2 \right)	\eqno(metricw)
$$
where $\rho\equiv\ln r$.  Moreover, the scalar field written as \Eq{ssphi}
appears irregular at $z$, but can be written in a regular form as
$$
        \phi(t,r) = r^{\iw} F(w),		\eqno(ssphiw)
$$
and $(Q(w),P(w))$ are derived in a regular way from $F(w)$.  In terms of the
new independent variable $w$, the equations of motion become
$$\eqalignno{
 {d\over dw}\pmatrix{Q\cr P}	&= \1{1-v^2}\pmatrix{\phv&-1\cr-1&\phv}
		\pmatrix{\bp&0\cr0&\bm}\pmatrix{Q\cr P},&(emw a)\cr
 {dv\over dw}	& = u-1,&(emw b)\cr
 {du\over dw}	& = {u+1\over v}\left[4\pi(\left|Q\right|^2 + \left|P\right|^2)
				-u\right],&(emw c)\cr
 {du\over dw}	& = -8\pi(u+1)\Re(Q^*P).&(emw d)\cr
}$$
{}From the last two equations, $u(w)$ can be expressed as
$$
 u(w) = 4\pi[\left|Q\right|^2 + \left|P\right|^2 +2v\Re(Q^*P)]. \eqno(getuw)
$$
The point $z=\infty$ is now marked by $v(w)=0$, and the system is clearly
regular at this point.  (The apparent pole at $v=0$ in \Eq{emw c} is
cancelled by a zero in the numerator, from \Eq{getuw}.)
The singular points $v=\pm1$ are, in contrast, true singular points of the
system,
and are, respectively, the past similarity horizon at $z=z_2$, and a new
future similarity horizon.

Integration of \Eqs{emw} is tantamount to integration of the field
equations in a timelike hypersurface $r=1$.  Due to self-similarity,
this in turn is tantamount to evolution in a timelike direction of
a whole spatial hypersurface $0\le r<\infty$, in the region outside
of any similarity horizons.

Numerically we proceed as follows.  Having found a solution for $0\le z\le
z_2$, we integrate it a little further in $z$ and then transform by
\Eqs{new} to the $w$ variables.  At this point $v$ is a little less than
1, and decreasing.  Then we integrate in $w$ (noting that the solution, if
it approaches $w=0$, it always continues smoothly through) until one of
two things happen.

The first alternative is that $u\rightarrow\infty$ at some point;  we
interpret such a point as an apparent horizon in spacetime, where radial
gauge coordinates (which we use throughout, see \Eq{metric}) break down.
Such an apparent horizon
probably means that the solution represents, not a choptuon, but
a black hole that is growing by self-similar accretion of scalar field.
We believe that such a solution, if continued further in a different
coordinate system, would always encounter a spacetime singularity within
the black hole, but we will not pursue this issue here.

The second alternative is that $v\rightarrow-1$ at some finite $w$.
We interpret this as a second, future similarity horizon.  The
complex choptuon behaves in this way.  What are
the proper boundary conditions at the future similarity horizon?
Above, at the past similarity horizon, we argued that $\phi$ should
remain regular because it arose from regular initial conditions,
and was not influenced by the spacetime singularity to its future.
That argument does not hold water here, because the future similarity
horizon is outside the domain of dependence of the initial data,
in particular, it is the Cauchy horizon of that domain.
Observers on the future similarity horizon will see data coming
from the singularity, and indeed it is a very interesting question
to ask what they will see.  Therefore no boundary condition at all
is enforced on $\phi$ at the future similarity horizon;  $(\phi,Q,P)$ may
be an arbitrary combination of regular and irregular solutions.
This is good, because we have already used up all our boundary conditions
at the axis and the past similarity horizon, and we would be embarrassed
to have to obey further boundary conditions.  However, we must still
decide how to continue the solution across the horizon, because the horizon
represents a singular point of \Eqs{emw}, and furthermore is a Cauchy
horizon in spacetime.  Evolution across a Cauchy horizon in classical
general relativity is never unique.  Quantum considerations, which might
fix the evolution in some way, are beyond the scope of this paper, though
later papers will treat it.  Therefore we will continue with a
conservative assumption about how to evolve.

Drop self-similarity for a moment, and look at the general boundary
conditions for a wave equation in curved spacetime.  A test wave function
$\phi$ is allowed to have discontinuity, or a discontinuity in some higher
derivative, across the characteristic surfaces of the wave equation,
which are the null hypersurfaces of spacetime.  However, if the
wave equation is coupled to gravity, a discontinuity in value of
$\phi$ will cause an infinite-mass singularity, and is therefore
forbidden: $T_{\mu\nu}\sim(\nabla\phi)^2\sim{\delta()}^2$.
However, a discontinuity in the first or higher derivative of $\phi$
is allowed.  In terms of the characteristic initial value problem,
the initial data for the wave equation on a characteristic hypersurface
is the value of $\phi$ itself, but does not include any derivatives
of $\phi$.

We now assume that spacetime continues to be self-similar to the
future of the future similarity horizon.
So turn now back to our self-similar equations for $\phi$.  The matching
rules for $\phi$ at the future similarity horizon are as follows.  The
wave function $\phi$ must be continuous across the horizon, to
forbid an infinite-mass singularity at the horizon.  The regular
part of $\phi$ represents incoming radiation crossing the horizon from
the regular region exterior to it.  In contrast, the irregular part of
$\phi$ consists of outgoing radiation originating very close to the
spacetime singularity, streaming along the similarity horizon.  In $w$
coordinates, the wave variables $(Q,P)$ of the irregular part of the
solution oscillate an infinite number of times while approaching the
future similarity horizon, and also die as a power law if $\kappa>0$.
However, if $\kappa=1$ the amplitude of the oscillations remains constant.
The irregular part of the solution is allowed to be discontinuous;
that is, the amplitude of the irregular part of $(Q,P)$ can be different
on the two sides of the horizon.  This causes at worst an (allowable) jump
in the stress-energy.

In general we expect to find that $(Q,P)$ contains both regular and
irregular parts at the future similarity horizon.  Then the regular
part must be continuous, but the irregular part is allowed to jump
arbitrarily.  The solution can still be fixed uniquely, if we are willing
to postulate that spacetime is smooth along the future time axis $t>0$,
$r=0$.  In that case, $\phi$ and the spacetime geometry must obey
boundary conditions at the future time axis, which are essentially the
same as the boundary conditions we have already imposed at the past
time axis.  Just counting degrees of freedom in the boundary conditions,
we expect this boundary condition to fix a unique solution (or perhaps a
discrete set of solutions).

However, at this point a numerical miracle happens at the future similarity
horizon.  We find that for the complex choptuon, $\phi$ is purely irregular
there, so that the regular piece of $\phi$ vanishes there, which entails
that $\kappa=1$ there (to an accuracy of $10^{-6}$), so that the future
similarity horizon carries initial data for flat spacetime with constant
$\phi$.  In particular the mass aspect vanishes on this null hypersurface.

Figure 3 displays the functions $Q(w)$, $P(w)$, $v(w)$, $u(w)$.  Note
that, at the future similarity horizon marked by $v=-1$, $u=0$, which
means the mass aspect vanishes there; this is the numerical miracle
discussed above.  As indicated in Figure 4, the scalar field is irregular
at the future similarity horizon, and oscillates and infinite number
of times.  A distant observer, measuring outgoing radiation in the
scalar field, would see such a signal coming from the threshold
gravitational collapse.

Consider again the evolution of the solution to the future of the future
similarity horizon.  In view of the boundary conditions, we are allowed
to choose an arbitrary irregular part just to the future of this horizon.
The choice of zero irregular part gives flat spacetime to the future of
this horizon.  Any other choice will give a negative mass in this region,
and will create a negative mass naked singularity along the future time axis
$t>0$, $r=0$.  Because we are attempting to evolve across a Cauchy horizon,
the choice is not obligatory.  However, the most sensible choice is clearly
the one that gives flat spacetime.  We therefore conclude that the
complex choptuon leaves behind it flat spacetime to the future of
the singular point at $(t,r)=(0,0)$, to an accuracy of 1 part in $10^6$.

\head{V. Discussion and Outlook}

In Figure 4, we show the interpretation arrived at for the complex
choptuon.  In region I, there is a collapsing ball of scalar field, bound
by its own self-gravity;  it acts as a near zone for scalar radiation.
The ball collapses toward the spacetime singularity at the origin, shown
as a single point.  As the ball collapses, scalar field is partially
trapped by spacetime curvature, but also continually leaks out of the
gravitationally bound region, across the past similarity horizon at
$v=+1$, and radiates outwards through regions IIa and IIb.  The boundary
between Region IIa and Region IIb is the surface $t=0$ in Choptuik
coordinates, or $z=\infty$.  However, this is merely a coordinate
singularity, and there is no physical boundary between the two regions;
they should be considered as a single Region II, which acts as a transition
zone between the near and far field for scalar radiation.  Spacetime is
curved in Region II, and in general some backscatter of outgoing radiation
to ingoing radiation is to be expected.  However, the numerical miracle
discussed in Sect.~IV indicates that the backscattered ingoing radiation
vanishes exactly at the future similarity horizon $v=-1$, for reasons that
we don't understand.  In any case, an irregular pulse of outgoing
radiation propagates outwards, with an infinite number of wavefronts
piling up at the future similarity horizon.  This is the nature of the
pulse that a distant observer sees.

This irregular pulse originates in the ``echoing" behavior of the collapse
in Region I.{\null}  Choptuik found the ``echoing" scale-factor
$$
	e^\Delta\approx30
$$
in his real choptuon, \Eqs{dss}.  In the complex choptuon the echoing
behavior is somewhat different, but the comparable scale for
the solution to recur is measured by $2\pi$ radians in the phase
oscillation of $\phi$, so that
$$
	e^{\Delta} = e^{2\pi/\omega} = 26.583086\pm0.000005,
$$
a value which clearly differs from Choptuik's.  Therefore, this critical
exponent $\Delta$ is not universal between the two solutions.

There is an exact self-similar solution for a collapsing pulse of real
scalar field $\phi$ in spherically symmetric general relativity\refto{TM,
Brady,ONT}, found independently by at least three groups.  It would not be
surprising to learn of further independent discoveries.  This solution ---
which we shall call the MBONT solution, after its several discovers ---
exhibits critical behavior of the black hole mass at
threshold\refto{Brady,ONT}.  However, the MBONT solution appears different
from the choptuons in important respects.  It does possess a past
similarity horizon; however the scalar field $\phi$ is not regular but
irregular there --- having a step in its derivative.  A closely related
issue is that the MBONT solution does not evolve from regular initial
conditions for gravitational collapse.  It remains to be seen whether or
not the MBONT solution is an attractor.  We shall return to this solution
in a future paper.

In this paper we have worked exactly at the threshold for black
hole formation, and have not addressed the critical behavior
of the black hole mass, \Eq{mbh}.  This critical behavior
can be determined by first order perturbation theory around
the complex choptuon, just as it can be addressed for the
Evans and Coleman hot-gas choptuons\refto{EC,epriv}.  Perturbation
theory can also determine whether or not the solution is an
attractor.  We will return to these issues in a future paper.

Region III, to the future of the singular point, is exactly flat in our
solution.  Since the future similarity horizon is a Cauchy horizon,
evolution cannot be unique; according to classical relativity, anything
could come out of the singular point.  However, it is remarkable that the
solution admits an evolution into exactly flat spacetime, and we have
argued that this is the preferred evolution.

Thus our solution provides one possible answer to the question: What is
the endstate of gravitational collapse at the threshold for black hole
production?  The endstate is an outgoing, irregular pulse of radiation
from the singular point in the wave zone , together with the most
conservative of all possible endstates in the near zone: Flat spacetime.

Further physical insight about this question must ultimately come from
quantum gravity or string theory, since strong curvatures comparable to
the Planck value occur near the singular point.

\head{Acknowledgements}

This research was supported in part by the National Science Foundation
under Grant Nos.~PHY89-04035 and PHY90-08502, and parts were carried
out at the Aspen Center for Physics.  We are especially grateful
to Matt Choptuik for providing some numerical truth by private communication,
and to Jim Horne for enlightening communications about this and related
problems.  We are grateful to J.~Traschen and D.~Kastor for helpful
conversations at ITP, and DME is grateful to C.~Evans, R.~Price,
K.~Grundlach, D.~Garfinkle, J.~Pullin and B.~Schmidt for helpful
conversations at Aspen.

\head{Appendix A}

In Table 1 we present all numerical solutions returned by our algorithm.
As explained in Sect.~IV, accuracy is lost near the past similarity
horizon $z=z_2$, due to decay of the irregular part of $\phi$ there,
\Eqs{irreg}.  Physically we would like the solution $\phi$ to be $C^\infty$
at $z=z_2$.  However, our numerical algorithm cannot distinguish
between degree of smoothness $C^3$ and $C^\infty$, because Taylor
expansions of the regular solution around $z_2$ were carried out only
through second order.  (If we had carried them to $n^{\rm th}$ order,
the algorithm could distinguish $C^n$ from $C^\infty$, but could not
distinguish $C^{n+1}$.)
In turn, the degree of smoothness of the irregular solution depends
on $\kappa$, \Eqs{irreg}, in such a way that our algorithm is reliable
for $\kappa\ltwid1$, but loses reliability for $\kappa\approx1/3$.
The solutions marked ``Reliable? No" in Table 1 all have $\kappa\approx1/3$
and therefore we do not know whether $\phi$ is irregular and $C^3$,
or regular and $C^\infty$, at $z=z_2$ for them.  The ``Interpretation''
depends on the behavior of the solution outside the past similarity
horizon at $z=z_2$.  Solutions marked ``Black Hole" encounter a
numerical singularity that appears to be an apparent horizon.  Such an
apparent horizon probably means that the solution represents, not a
choptuon, but a black hole that is growing by self-similar accretion
of scalar field.  Solutions marked "Choptuon" encounter no such
apparent horizon, but do encounter a future similarity horizon.

\goodbreak\vskip 0.5truein{\narrower
	\noindent TABLE 1.\quad Numerical solutions found.  Solution is
	started at $z=0$ with values $\omega$ and $P$, in \Eqs{data}.  The
	past similarity horizon appears at $z=z_2$.  All solutions marked
	``Reliable? No" have $\kappa\approx1/3$ there, and hence may have
	lost accuracy.\par}

\bigskip
{\singlespace\halign
{\hskip0.5in#\quad\hfil&#\quad\hfil&#\quad\hfil&#\quad\hfil&#\quad\hfil\cr
$\omega$	&$P$	&$z_2$	&Reliable? &{\sl Interpretation}\cr
\multispan5\hskip0.5in\leaders\hrule\hfill\cr
0.0      &0.16286  &1.10668	&Yes	&Robertson-Walker solution\cr
0.0      &0.21412  &1.24793	&No	&Black Hole?\cr
0.15265  &0.23468  &1.35664	&No	&Black Hole?\cr
0.29825  &0.24502  &1.40583	&No	&Black Hole?\cr
0.43076  &0.25919  &1.47793	&No	&Black Hole?\cr
0.55509  &0.27670  &1.57717	&No	&Black Hole?\cr
0.68307  &0.29923  &1.72316	&No	&Black Hole?\cr
0.81757  &0.32998  &1.97453	&No	&Black Hole?\cr
0.94825  &0.37269  &2.49240	&No	&Choptuon?\cr
1.04980  &0.44740  &4.56352	&No	&Choptuon?\cr
1.91544  &0.67217  &5.00354	&Yes	&Complex Choptuon;  see Sect.IV\cr
}}

\references
\oneandathirdspace

\refis{Chop} M.W. Choptuik,
\prl 70, 9-12, 1993.

\refis{DC} See also D. Christodoulou, \journal Comm. Math. Phys.,
105, 337-361, 1986;
{\bf106}, 587-621 (1986); {\bf109}, 591-611, (1987); {\bf109}, 613-647, (1987).

\refis{EC} C.R. Evans, \& J.S. Coleman,
\prl 72, 1782-1785, 1994.

\refis {AE} A.M. Abrahams \& C.R. Evans,
\prl 70, 2980-2983, 1993.

\refis{Brady} P.R. Brady,
\journal Classical and Quantum Gravity, 11, 1255-1260, 1994.

\refis{ONT} Y. Oshiro, K. Nakamura \& A. Tomimatsu,
\journal Progr. Theo. Phys., 91, 1265-1270, 1994.

\refis{ST} A. Strominger \& L. Thorlacius,
\prl 72, 1584-1587, 1994.

\refis{JT} J. Traschen,  {\sl Discrete Self-Similarity and Critical Point
Behavior in Fluctuations about Extremal Black Holes,} preprint gr-qc/9403016,
1994.

\refis{TM} Maithreyan, T., unpublished Ph.D. Thesis, Boston University, 1984.

\refis{DG} Garfinkle, D. {\sl Choptuik scaling in null coordinates,}
preprint gr-qc/9412008, 1994.

\refis{cpriv} M. Choptuik, private communication (1994).

\refis{epriv} C. Evans, private communication (1994).

\refis{DE} D.M. Eardley,
\journal Comm. Math. Phys., 37, 287, 1974.

\refis{Bev} B.K. Berger, \jmp 17, 1268, 1976.

\refis{PenCC} Penrose, R. 1979.  \rm  {\it In} General Relativity, an
Einstein Centennary Survey.  S.W. Hawking and W. Israel, Eds.:581-638.
Cambridge University Press.  Cambridge.

\refis{NR} W.H. Press, B.P. Flannery, S.A. Teukolsky, and W.T. Vetterling,
{\it Numerical Recipes} (Cambridge University Press, Cambridge, 1986).

\refis{HE} S.W. Hawking \& G.F.R. Ellis, {\it The Large Scale Structure of
Space-Time} (Cambridge University Press, Cambridge, 1973).

\refis{EIMM} D. Eardley, J. Isenberg, J. Marsden \& V. Moncrief,
{\sl Homothetic and Conformal Symmetries of Solutions to Einstein's
Equations,}
\journal Comm. Math. Phys.,  106, 137--158, 1986.

\endreferences

\figurecaptions

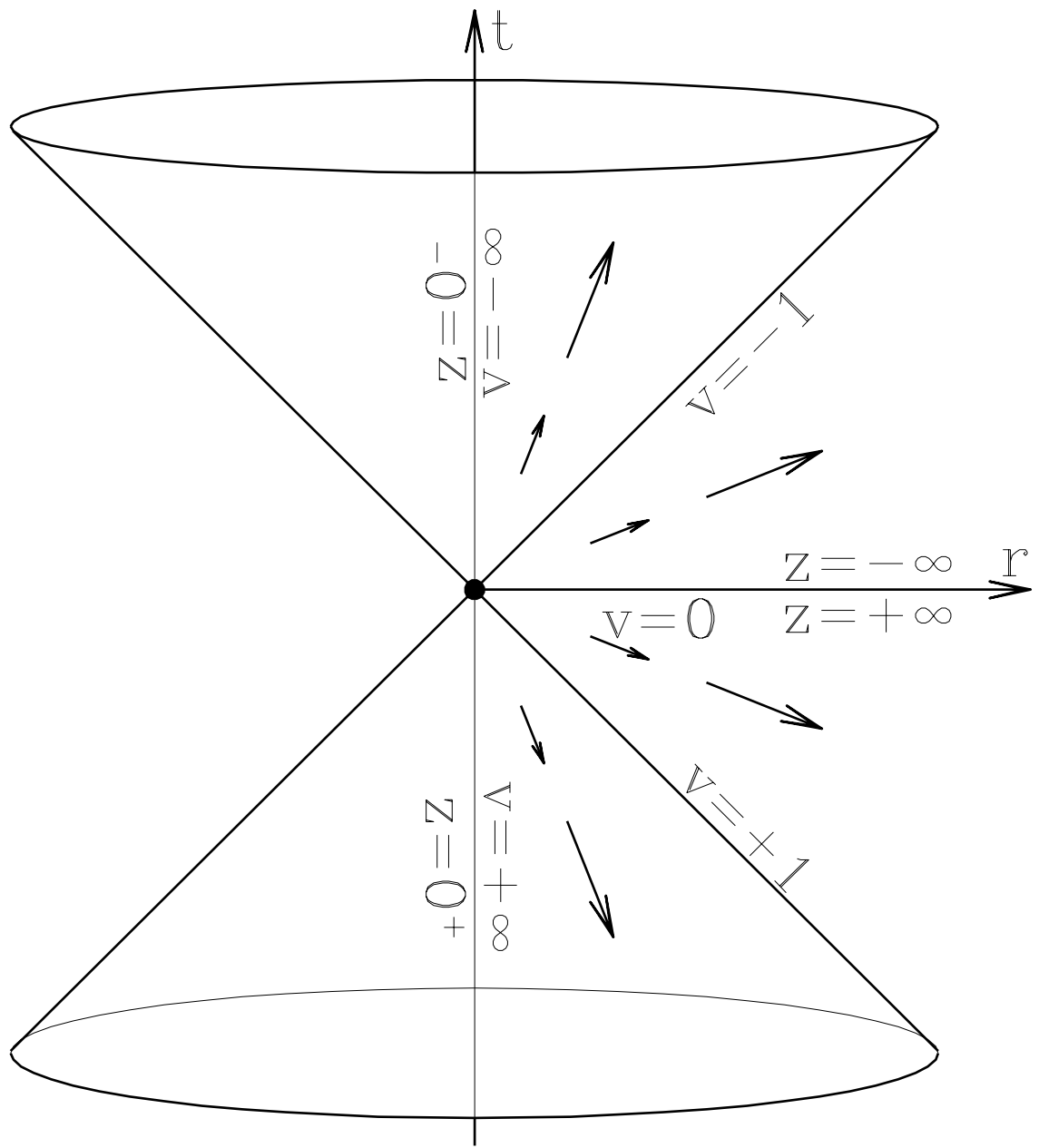
Figure 1.  Coordinate systems in Minkowski spacetime.  Arrows show the
homothetic killing vector field $\xi\equiv t\partial_t+r\partial_r$.
Shown also are the past and future light cones of the origin;  in the
terminology of this paper, these light cones are past and future
similarity horizons, where $\xi$ becomes null.  For coordinates $z$
and $v$, see Sect.~III.

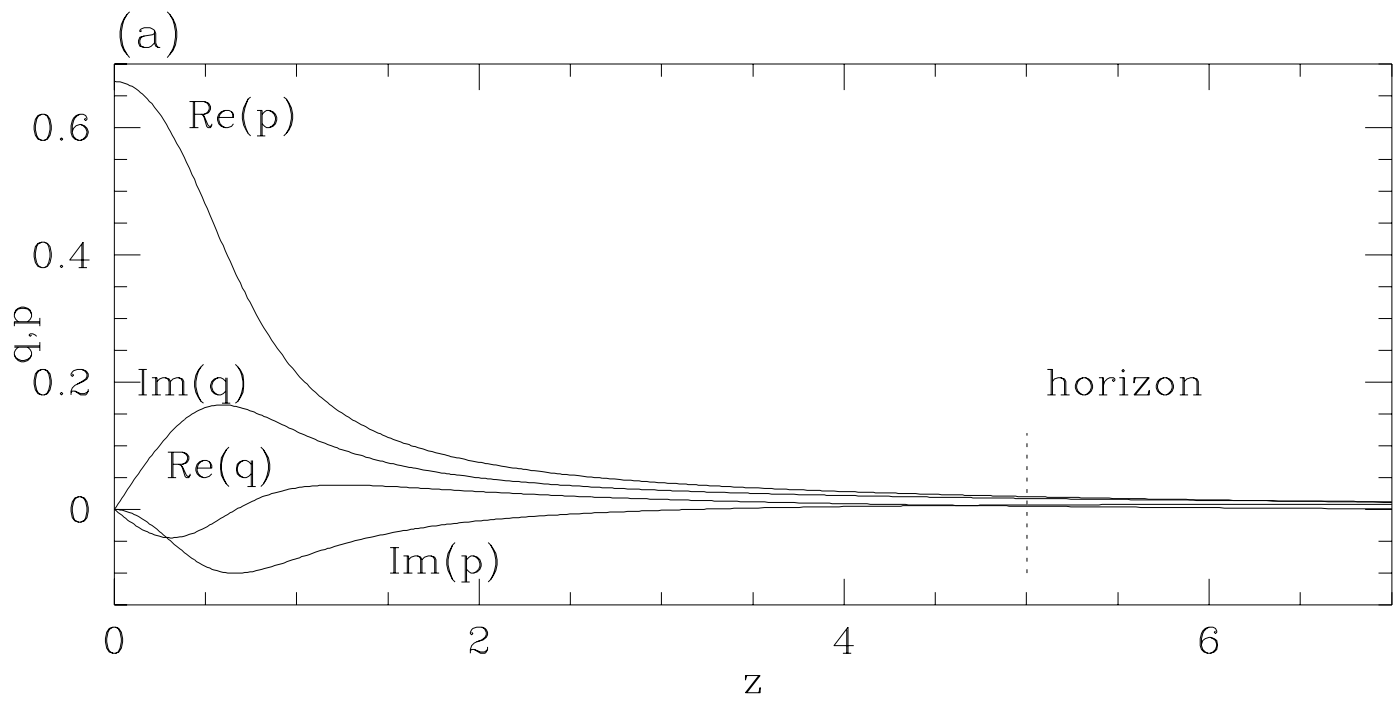
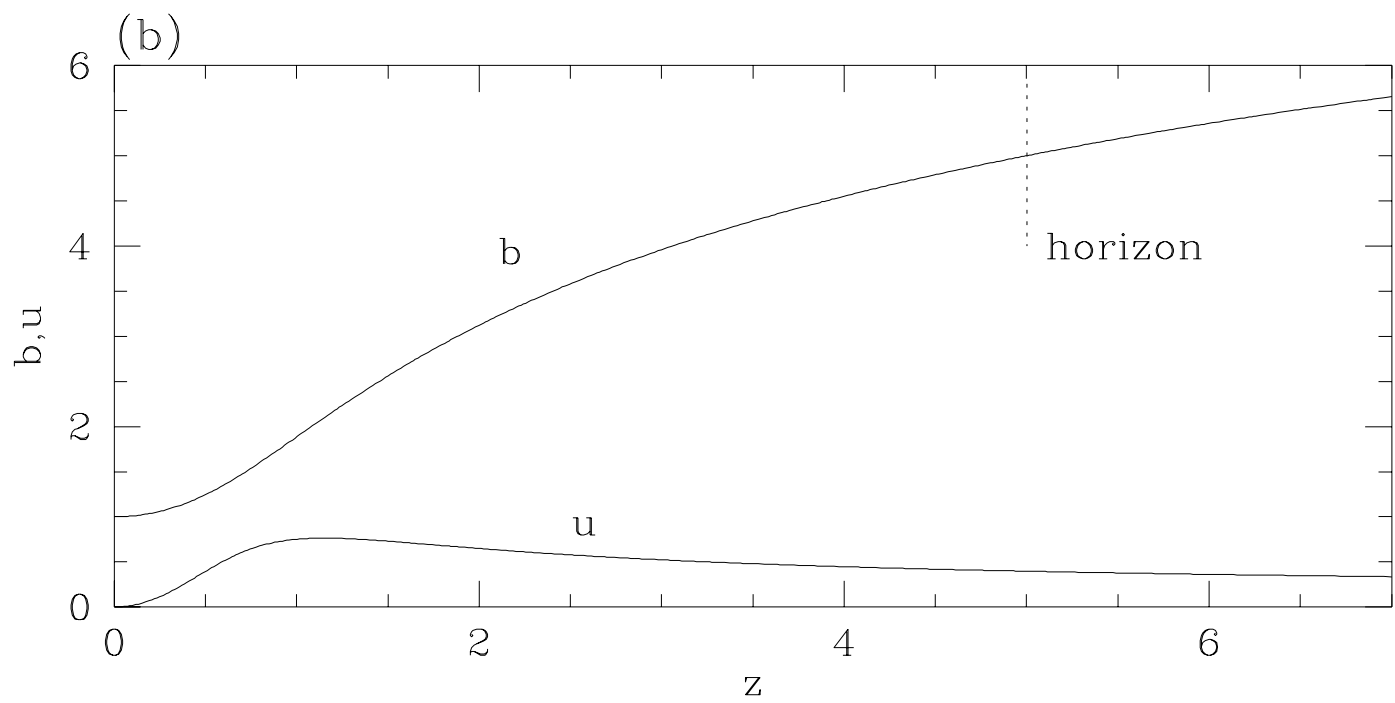
Figure 2.  Behavior of the complex choptuon, the solution for
$\omega=1.9154446$, over the domain $0\le z\le10$.  The past similarity
horizon is located at $z_2=5.0035380$, and $\omega$ was determined by
demanding regularity there.  (a)  The complex functions $q(z)$ and $p(z)$
which represent the scalar field $\phi$.  (b)  The metric functions $b(z)$
and $u(z)$.

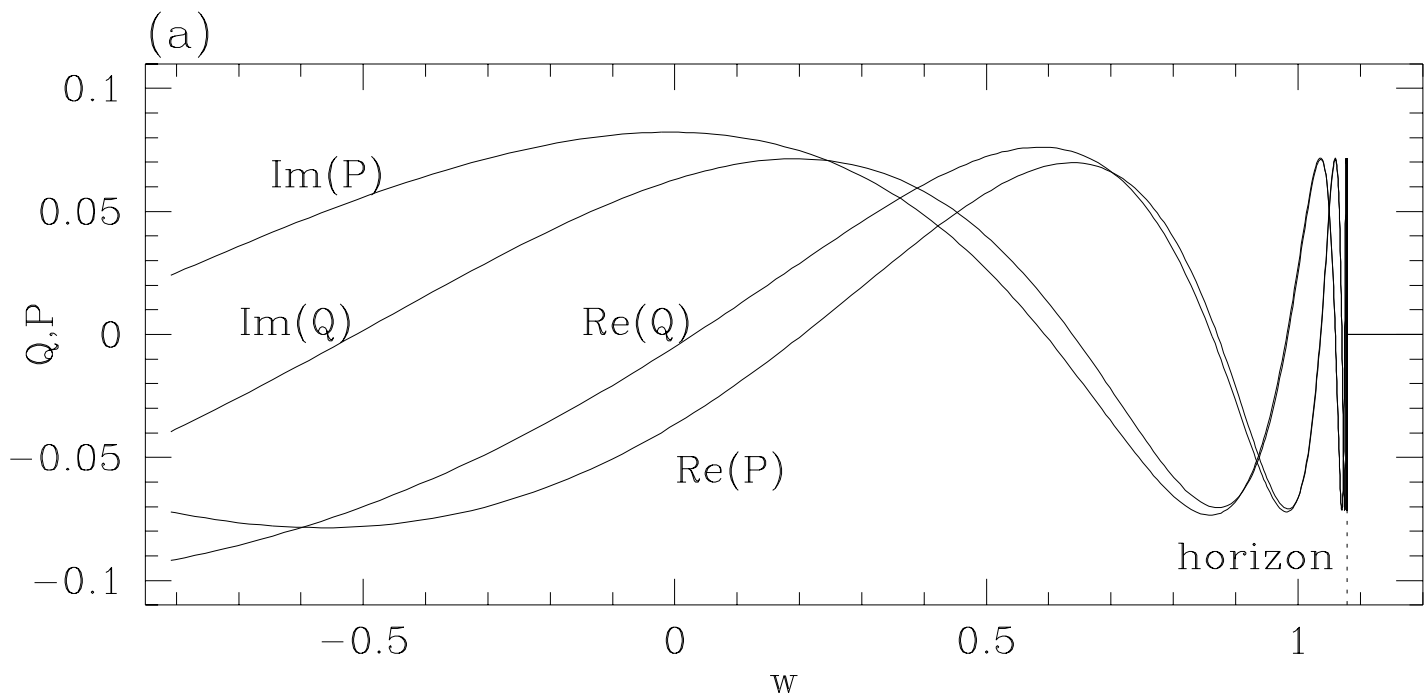
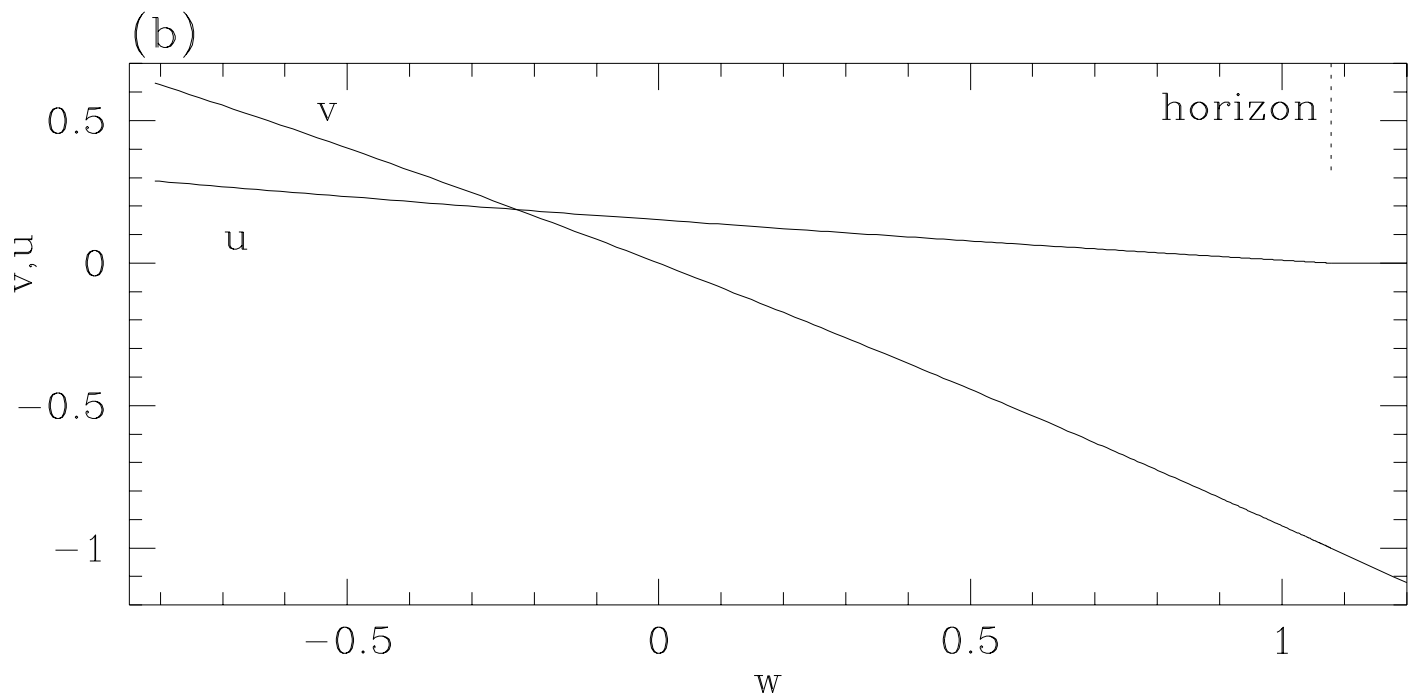
Figure 3.  Behavior of the complex choptuon, the solution for
$\omega=1.9154446$, over the domain $-0.8\le w\le 10.$  The future
similarity horizon is located at the value $z_4$ of $z$ where $v(z)=-1$.
(a)  The complex functions $Q(w)$, $P(w)$ which represent the scalar field
$\phi$.  They oscillate infinitely many times in approaching the future
similarity horizon. (b)  The metric functions $v(z)$ and $u(z)$.  Note
that $u(z_4)=0$ to numerical accuracy, showing that the solution matches
onto flat spacetime at the future similarity horizon.

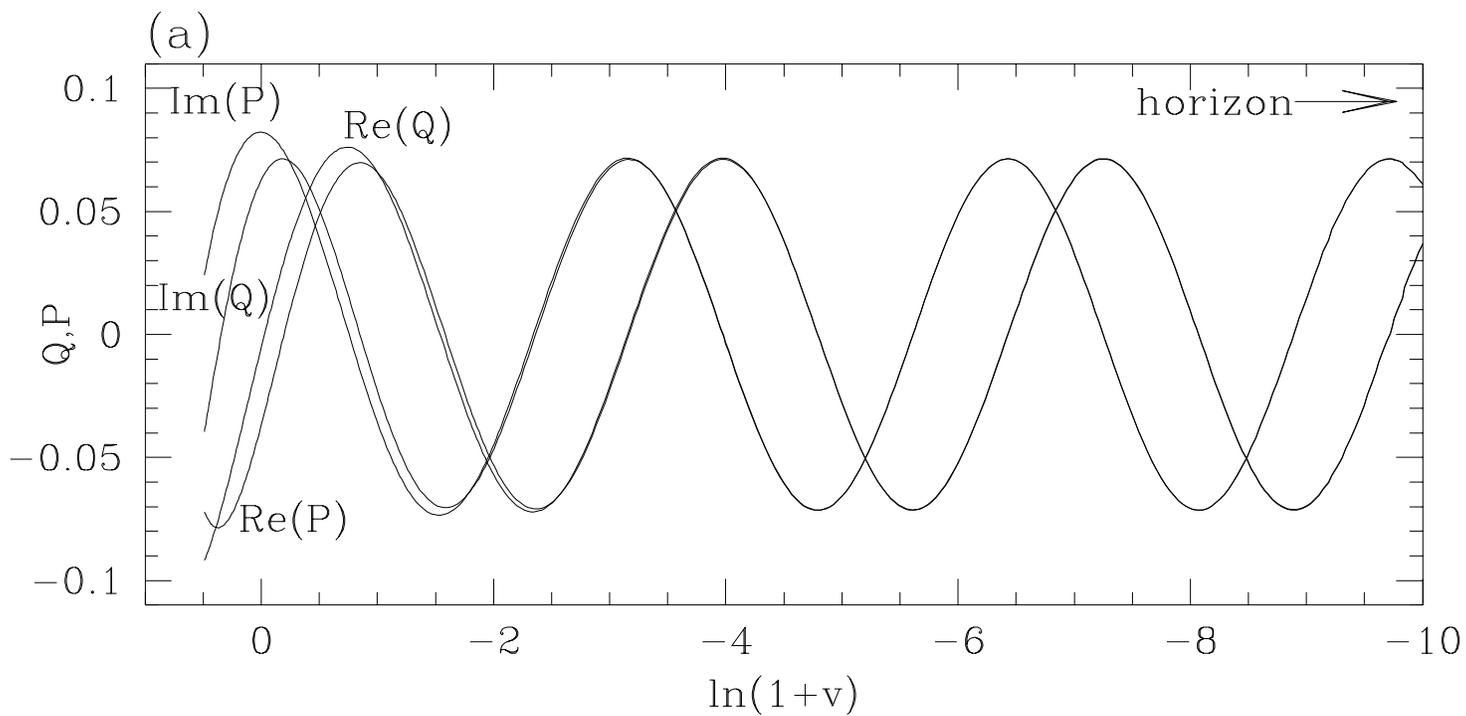
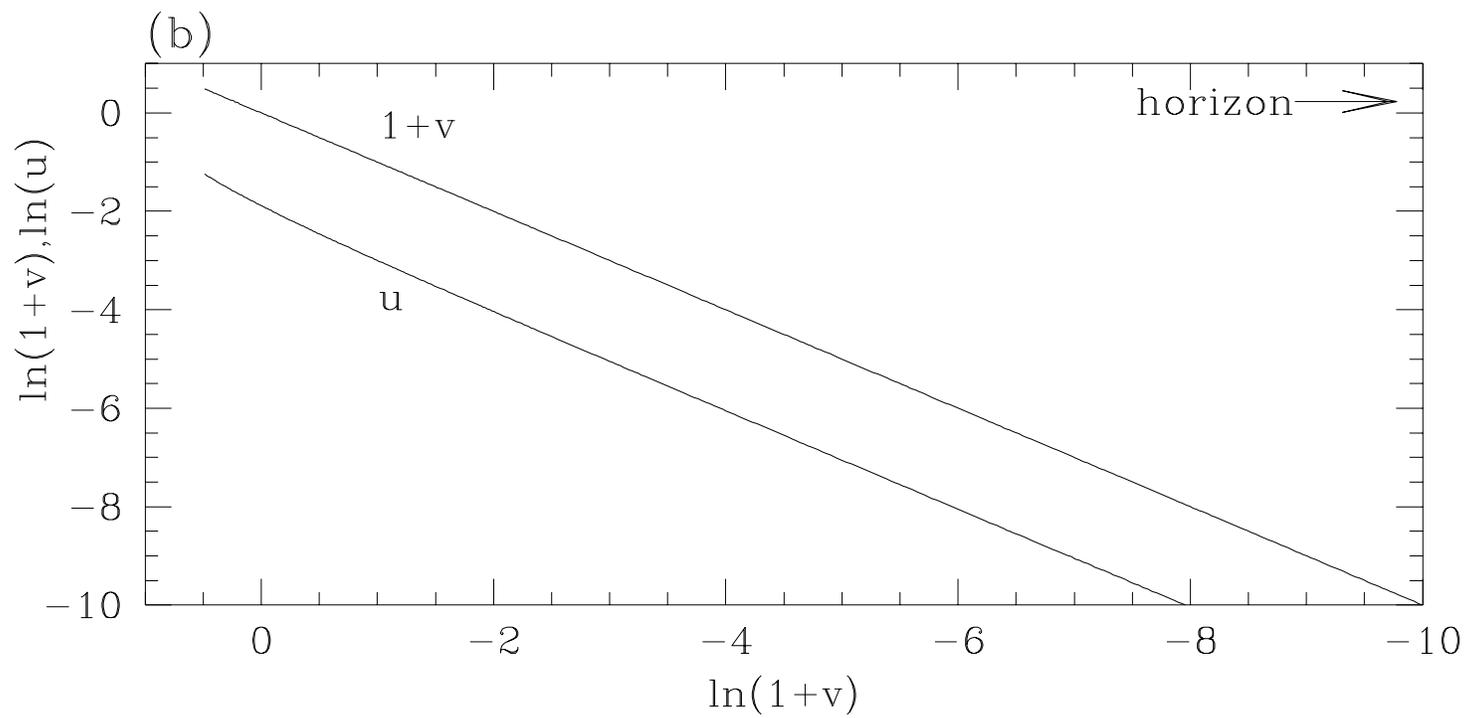
Figure 4.  Same as Fig.~3, except that the horizontal axis is now
logarithmic, and shows $\ln(1+v)$. (a) The complex functions $Q(w)$,
$P(w)$ which represent the scalar field $\phi$.  They oscillate infinitely
many times in approaching the future similarity horizon. (b)  The metric
functions $v(z)$ and $u(z)$, plotted as $\ln(1+v)$ and $\ln(u)$, both
of which approach 0 on the future similarity horizon.

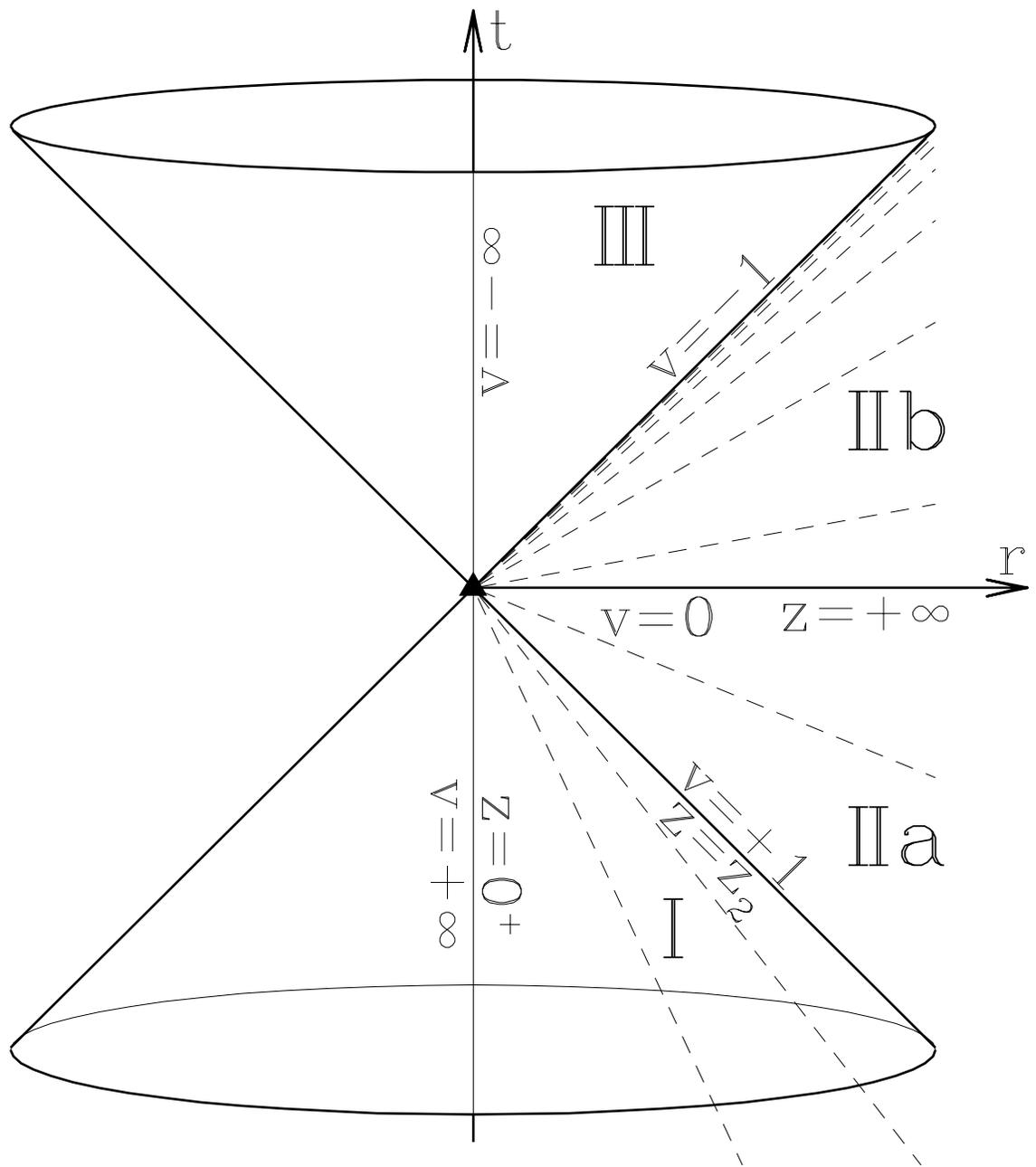
Figure 5.  Interpretation of the complex choptuon.  The similarity
horizons lie at $v=+1$ (past) and $v=-1$ (future).  Dotted lines show
peaks and valleys of the scalar field ($q$ for $v<-1$, $Q$ for $v>-1$), to
illustrate its oscillations.  Region I is a collapsing sphere of
gravitationally bound scalar field; in Regions IIa and IIb this blends
smoothly into an outgoing scalar wave.  The outgoing scalar wave
oscillates infinitely many times approaching the future similarity horizon
at $v=-1$.   Region III is flat.

\endit